\documentclass{iau-JDSS}
\usepackage{graphicx}

\title[Evolution of NSCs] 
{Formation, Growth, and Destruction of Nuclear Star Clusters}

\author[T. B\"oker]   
{Torsten B\"oker }

\affiliation{European Space Agency, Keplerlaan 1, 2200AG Noordwijk, 
Netherlands \break email: tboeker@rssd.esa.int} 

\pubyear{2009}
\volume{Volume 15}  
\pagerange{119--126}
\date{?? and in revised form ??}
\jname{Highlights of Astronomy, Volume 14}
\editors{Ian F Corbett, ed.}
\begin{document}

\maketitle

\begin{abstract}
This talk is an attempt to combine recent insights into the nature of the nuclear star clusters in galaxies 
of various morphologies into a coherent (albeit simplistic) picture for their formation, growth, 
and eventual destruction.
\keywords{galaxies: nuclei; galaxies: star clusters}
\end{abstract}

\firstsection 
\section{Formation and Growth of Nuclear Star Clusters}
The structural properties, masses, and stellar populations of nuclear star clusters (NSCs) have been extensively 
discussed elsewhere, both for late-type spirals \cite[(B\"oker et al. 2004, Walcher et al. 2005, 2006)]{boeker04} 
as well as for spheroidal galaxies \cite[(C\^ot\'e et al. 2006)]{cote06}. A key finding from these studies 
is that, in general, the star formation histories of NSCs are long and complex, but {\bf only} NSCs in late-type 
spirals also contain a young (few hundred Myrs) stellar population. The NSCs in spheroidal galaxies, 
on the other hand, generally have luminosity-weighted ages of many Gyrs, but are on average 
3.5\,Gyrs younger than the bodies of their host galaxies \cite[(Paudel 2010)]{paudel10}.

This implies that in the present-day universe, any growth mechanism that causes the ``rejuvenation'' 
of NSCs occurs only in (gas-rich) disk galaxies. The NSCs in (gas-poor) spheroidals, on the other
hand, must have evolved passively for at least a few Gyrs. If they have acquired stellar mass over 
this timescale, it can only have occurred via the accretion of evolved stars, e.g. through the merging of 
globular clusters onto the NSC.

The merging of star clusters is indeed one of the proposed mechanisms for the build-up of NSCs
\cite[(Capuzzo-Dolcetta \& Miocchi 2008)]{capuzzo08}. At least in disk galaxies, however, this is unlikely 
to be the whole story, because
\cite{hartmann11} show that cluster infall alone cannot explain the dynamical properties of 
the NSC in the nearby edge-on disk NGC\,4244 . They conclude that at least 50\% of its 
mass must have been produced by in-situ star formation. The star formation is the result
of gas infall into the central few pc, which can be caused by a number of mechanisms such as 
bar-induced torques \cite[(Schinnerer et al. 2006)]{schinnerer06}, compressive tidal forces 
\cite[(Emsellem \& van der Ven 2008)]{emsellem08}, or the magneto-rotational instability in 
galaxy disks \cite[(Milosavljevic 2004)]{milos04}.

In any case, it is important to keep in mind that the {\it formation} of NSCs and their subsequent {\it growth} do 
not necessarily have to be governed  by the same mechanism, and that more than one mechanism can 
contribute to the evolution of an NSC after its formation. 

\section{SMBHs and the Destruction of Nuclear Star Clusters}\label{sec:destr}
The co-evolution of NSCs and central super-massive black holes (SMBHs) is currently a very active field 
of research, triggered by the realization that both types of a central massive object (CMO) can co-exist 
at the low end of the SMBH mass range \cite[($M_{\rm BH} < 10^9{\rm M_{\odot}}$, Seth et al. 2008)]{seth08}. 
Unfortunately, the exact mechanism for SMBH formation, and whether or not an NSC is {\it necessary} to
form an SMBH remain unclear for now.

However, there appears to be a transition in what type of CMO dominates: at very low CMO masses,
SMBHs are hard to identify \cite[(Satyapal et al. 2009)]{satyapal09}, and their mass is usually less than that of the NSC. 
In galaxies hosting SMBHs with masses above $\approx 10^{10}{\rm M_{\odot}}$, on the other hand, NSCs are usually 
not observed \cite[(Graham \& Spitler 2009)]{graham09}, suggesting that the most massive SMBHs have destroyed 
their host NSCs. A similar transition is also evident in the surface brightness profiles of spheroidal galaxies, which 
smoothly transition from a pronounced light excess in the central few pc of low-mass systems to a clear light deficit
in high mass system \cite[(C\^ot\'e et al. 2007)]{cote07}. This gradual change may be identified with a decreasing
prominence of the NSC as the mass of the host spheroid grows. 

The mass ratio between SMBH and NSC may well be imprinted early-on in the galaxy's life by their 
mutual feedback during the ``competitive accretion'' phase, as proposed by \cite{nayakshin09}. However,
it is also tempting to speculate that, as the evolution of the host galaxy progresses, the SMBH grows 
in mass at the expense of the NSC. For example, the merger of two BHs has been shown to efficiently 
destroy the surrounding NSC(s) via loss-cone depletion \cite[(e.g. Merritt 2006)]{merritt06}. 
While many details of NSC evolution remain to be addressed, the general picture outlined in Figure 1 
appears to be consistent with what we currently know.

\begin{figure}
\begin{minipage}[h]{0.5\textwidth}
\includegraphics[height=2.2in]{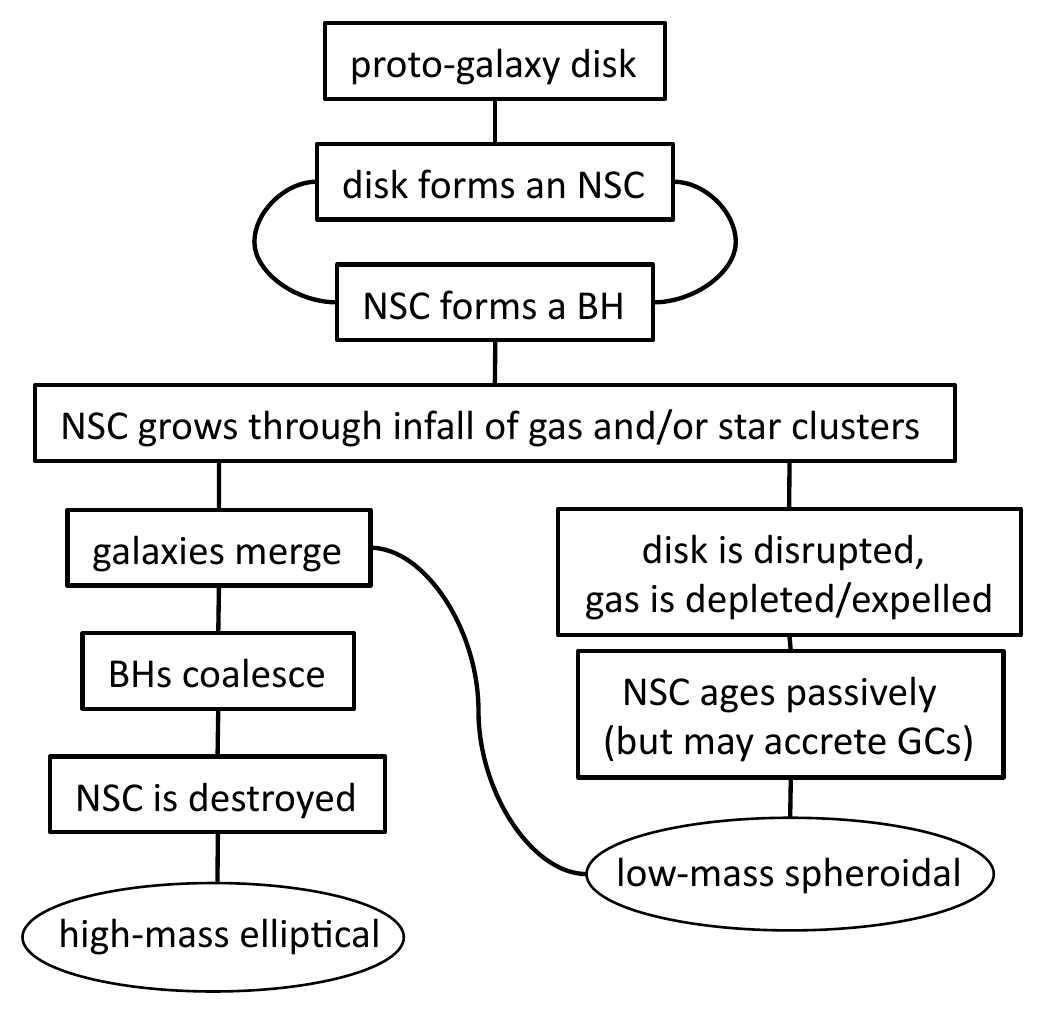} 
\end{minipage}\begin{minipage}[h]{0.5\textwidth}
{{\bf Figure\,1}  Possible evolutionary path of an NSC that forms early in the
life of a gas-rich, disk-dominated ``proto''-galaxy. As long as gas is available for 
infall into the center, the NSC (and any SMBH it may harbor) will grow. 
Once the gas is depleted or expelled (e.g. by the disruption of the galaxy disk 
through harassment or a close encounter), the CMO growth stops, and it will only evolve 
passively from then on. The host galaxy may progress along the merger tree, and its SMBH may
well merge with the SMBHs of the merger partners, destroying their hosts' NSCs in the process.}
\end{minipage}
\end{figure}

\end{document}